\documentstyle[prl,epsf,aps]{revtex}
 \newcommand {\bi} {\bibitem}
 \newcommand {\be} {\begin{equation}}
\newcommand {\bea} {\begin{eqnarray} \nonumber }
\newcommand {\ee} {\end{equation}}
\newcommand {\eea} {\end{eqnarray}}
 \newcommand {\eps} {\epsilon}
 \newcommand {\si} {\sigma}

\newcommand {\la} {\lambda}

\newcommand {\R} {{\cal R}}

\newcommand {\lan} {\langle}
\newcommand {\ran} {\rangle}

\def \form#1 {eq. (\ref{#1}) }
\def \parziale#1#2  {{\partial {#1} \over \partial {#2}}}

\begin{document}
\twocolumn
[\hsize\textwidth\columnwidth\hsize\csname@twocolumnfalse\endcsname

\title{A divergent correlation length in off-equilibrium glasses}
\author{
Giorgio
Parisi}
\address{ Dipartimento di Fisica, Universit\`a {\em La  Sapienza},\\ 
INFN Sezione di Roma I \\
Piazzale Aldo Moro, Rome 00185}
\maketitle

\begin{abstract}
In off-equilibrium dynamics we  define a dynamical correlation length which is proportional to 
the 
size of the region in which the atoms move in a correlated way.  General arguments  
indicate that this dynamical correlation length diverges at large times in the glassy phase.  
Numerical simulations for binary mixtures point toward the correctness of this prediction.
\end{abstract}
\vskip.3cm
]
\narrowtext 

The glass transition has many puzzling features, one being the presence of a divergent time scale 
without a corresponding divergence of the correlation length as measured by considering {\sl 
equilibrium} correlations.  In this letter we will show that a divergent correlation length may be 
identified in the glassy phase, below the glass transition point, if we study the {\sl 
off-equilibrium} correlation functions.

The basic idea is quite simple.  In a glass atoms are usually frozen in some position.  If they try 
to move, the other atoms push them in the original position.  This cage effect is at the origine of 
the glass transition.  At low temperatures, but still in the liquid phase, the viscosity is finite 
(albeit very large) and some movements are still allowed.  In many  theoretical approaches 
\cite{vetro} these movements correspond to the formation of relatively large fluidified domains.  
All the particles in the domain move of a certain amount and the radius of the domain diverges when 
we approach the transition.  The Vogel-Fulcher law may be obtained if the volume of these fluidified 
domains diverges at $(T-T_{K})^{-1}$ and the process is controlled by the energy activation which 
should be proportional to the volume.  The existence of processes involving a large number of 
particles in the liquid phase near the glass transition has been recently shown in
\cite{KG}.

When we decrease the temperature, the number of particles involved increases; these processes 
involve the crossing of higher and higher barriers and they are extremely hard to observe in 
equilibrium simulations \cite{LAPA}.  Here we study similar processes which are present when we 
quench the system from an high temperature to a temperature well below $T_{g}$
\cite{PAAGE1,KB,BAPA}.  The dynamics of systems going toward equilibrium has been the subject of a 
wide theoretical interests
\cite{OFF}.
In the high temperature phase the energy usually approaches equilibrium exponentially fast.  On 
the contrary in the low temperature phase the energy has corrections which decrease as a power of 
of 
time:
\be
E(t)=E_{\infty}+A\ t^{-\la(T)}.\label{ENE}
\ee
The dependence of $\la(T)$ on the temperature is quite instructive.  There are two simple 
possibilities.  In the first one, the approach to equilibrium is dominated by entropic barriers and 
the value of the exponent $\la(T)$ is roughly independent from the the temperature.  In the second 
case the approach to equilibrium is dominated by energy barriers: the value of $\la(T)$ strongly 
depends on the temperature and vanishes linearly with the temperature.  Depending on the nature of 
the system, the approach to equilibrium may be dominated by entropic or by energetic barriers.  Two 
well studied cases, which correspond to the two different behaviours, are the spinodal decomposition
\cite{BRAY} and the spin glasses \cite{MPRR}.

If coming from the high temperature phase we quench a ferromagnetic Ising system at temperatures 
much smaller than the critical one, we find bubbles in which the spins have a similar value and the 
radius of these bubbles increases at $t^{1/z}$ with $z=2$ \cite{BRAY}.  The radius of these bubbles 
coincide with our dynamical correlation length.  The persistence time of a bubble of size $R$ is 
proportional to $R^{z}$: a bubble which is present at time $t$ after the quench takes a time $O(t)$ to 
disappear.  Here $\la(T)=1/2=1/z$.  All exponents are temperature independent.  The spin-spin 
correlation $C$ satisfies the scaling form
\be
C(t,t_{w},r)\equiv\lan\si(x,t_{w})\si(y,t)\ran\approx S({t\over t_{w}},{r\over 
R(t_{w})})\label{FERRO}
\ee
where $|x-y|=r$ and $R(t_{w})\propto t_{w}^{1/z}$.  Eq.  (\ref{FERRO}) implies aging \cite{B}: at fixed 
$r$,
at $t$ and $t_{w}$ {\sl both} large, the correlation function depends only on the ratio $t/t_{w}$.

An opposite behaviour is seen in spin glasses.  The local magnetizations ($m(i)\equiv \lan 
\si(i)\ran$, $i$ being a point of the lattice) fluctuate from point to point with zero average (at
zero external magnetic field).  In order to observe an interesting behaviour it is convenient to 
consider two copies of the same system ($\si$ and $\tau$) and to introduce the overlap 
$q(i)=\si(i)\tau(i)$.  The quantity $q$ plays the same role of the magnetization in ferromagnets: 
the susceptibility associated to $q$ diverges at the critical point; the statistical expectation 
value of $qi(i)$ is equal to $m(i)^{2}$ and it may different from zero only below the transition.  The 
dynamical correlation function of $q$ gives important information on the dynamics.  In this case  
the theoretical understanding is not as good as in the case of ferromagnets; however 
intensive numerical simulations \cite{MPRR} show that (in three dimensions) the dynamic correlation 
length diverges as a power of the time and that the $q$-correlation function has a behaviour which 
is qualitatively similar to that of a ferromagnet:
\be
C_{q}(t,t_{w},x)\equiv\lan q(x,t_{w})q(y,t)\ran\approx r^{-\omega}S_{q}({t\over t_{w}},{r\over 
R(t_{w})}),
\ee
where also here $R(t_{w})\propto t_{w}^{1/z}$.  The main differences among spin glasses and 
ferromagnets are: (a) the presence of the prefactor $r^{-\omega}$ ($\omega
\approx .5$); (b) the value of the exponent $z,$ which is approximately given $z_{c}T/T_{c}$ 
($z_{c}\approx 7$ and $T_{c}$ is the critical temperature).  Also the relation among $z(T)$ and
$\la(T)$ is different from that of ferromagnets: the energy approaches the equilibrium value  
with 
an exponent $\la(T)
\approx 2.5 z(T)^{-1}$, which it is proportional to the temperature.

Summarizing ferromagnets and spin glasses show a clear cut behaviour characterized respectively by 
entropic and energetic barriers.  Which are the theoretical expectations for structural glasses?  
We 
can answer to this question if we assume that the mean field for some generalized spin glasses can 
be applied also to structural glasses \cite{KWT,parisi}.  This conjecture can be considered as a 
rationalisation of the Gibbs Di Marzio approach; as a byproduct it implies the correctness of the 
mode coupling theory in an appropriate time-temperature window \cite{BCKM}.

In this approach the decay of the energy has a rather interesting behaviour.  In the mean field 
theory there are two time windows.  In the first one, at relative short times, the approach to 
equilibrium is dominated by entropic barriers and the value of the exponent $\la(T)$ is roughly 
independent from the the temperature.  At larger times the approach to equilibrium is dominated by 
crossing of energy barriers; in this regime the time evolution strongly depends on the 
temperature.  
Numerical simulations (done both with Hamiltonian and dissipative dynamics
\cite{KB,BAPA}) for structural glasses are consistent with this picture.  Here we will
show that in the first (entropy dominated) regime there is dynamic correlation length which 
increases as power of time.

We present the results of a simulation for binary fluids.  We consider a mixture of soft particles 
of 
different sizes.  Half of the particles are of type $A$, half of type $B$ and the interaction among 
the particles is given by the Hamiltonian:
\begin{equation}
H=\sum_{{i<k}} ((d(i)+d(k))^{12}|{\bf x}_{i}-{\bf x}_{k}|^{-12},\label{HAMI}
\label{HAMILTONIAN}
\end{equation}
where the radius ($d$) depends on the type of particles.  This model has been carefully studied 
in 
the past \cite{HANSEN,PAAGE1,PAAGE3}.  The choice $d_{B}/d_{A}=1.2$ strongly inhibits 
crystallisation and the system goes into a glassy phase when it is cooled.  Using the same 
conventions of  previous investigators we consider particles of average radius $1$ at unit 
density.  It is usual to introduce the quantity $\Gamma
\equiv \beta^{4}$.  For quenching from $T=\infty$ the glass transition is known to happen around
$\Gamma_c=1.45$
\cite{HANSEN}. For computational reasons we 
have slightly modified this model by introducing a cutoff, i.e.  we have put to zero the 
interaction 
when $|x_{i}-x_{k}|^{2}>3$.  The differences in various thermodynamic quantities with the original 
model are of the order of 1\%.

\begin{figure}[htbp]
  \epsfxsize=250pt\epsffile{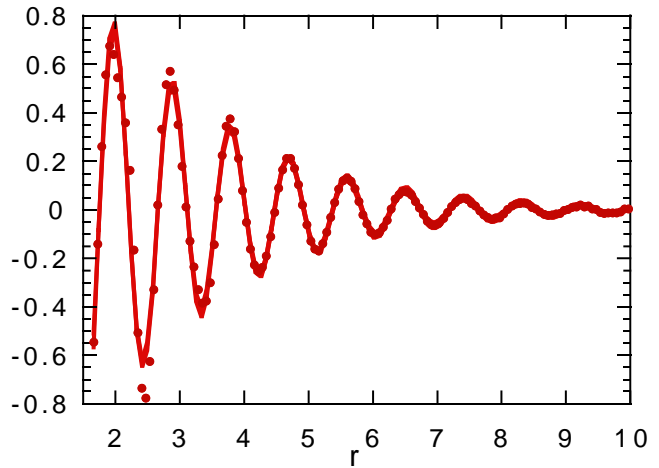}
\caption{The function $r(g(r)-1)$ versus $r$, the values of the fit parameters (according 
to eq.  (5)) are for $a=0$: $r_{g}=1.29$, $d=.89$, $\phi=-1.15$.  Superimposed there is 
an indistinguishable fit with $a=1$.}
\label{G}
\end{figure}

Our simulation are done using a Monte Carlo algorithm, which is more easy to deal with than 
molecular dynamics, if we change the temperature in an abrupt way.  Each particle is shifted by a 
random amount at each step, and the size of the shift is fixed by the condition that the average 
acceptance rate is about .4.  Particles are placed in a cubic box with periodic boundary conditions 
and at the end of each Monte Carlo sweep all the particles are shifted of the same vector in order 
to keep the center of mass fixed \cite{LAPA}.  We have done simulations for system of many sizes.  
Here we present the results for systems with 27000 particles, which correspond to a box of size 
30.  
We need to use large systems in order to avoid finite volume effects
\cite{HKBA}
(the size of the system must be much larger than the dynamical correlation length we  
study).  
The value of $\Gamma$ is 1.8, which is deep in the glassy region (it corresponds to a temperature 
three times smaller that the transition temperature \cite{HANSEN}).

Let us recall the results for the density-density correlation function ($g(r)$) at equal time at 
equilibrium after a rapid quench. The correlation function is shown in  fig.(\ref{G}) at distances greater than $1.6$; in this region it can be fitted as
\be g(r)=1+A r^{-a} \exp(-r/R_{g})\sin(2\pi r/d+\phi),\label{FIT1}\ee
which correspond to a pair of complex singularities in momentum space (simple poles for $a=1$). The 
fitted value of $R_{g}$ strongly depends on $a$; we find $r_{g}=$ 1.29 and 2.13 respectively for 
$a=0$ and $a=1$.  A fit done only at large distance (in the region $r>4$) is better and gives 
similar values of $R_{g}$ (1.88 and 1.41).  The value of the correlation length often strongly 
depends on the power in the prefactor and, unless one has extremely good data, it is difficult to 
determine {\sl both} the power of the prefactor and the rate of the exponential decay.

\begin{figure}[htbp]
\epsfxsize=250pt\epsffile{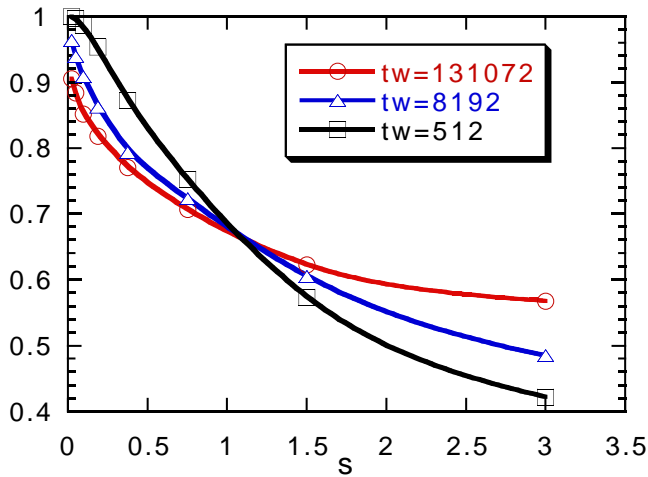}
\caption{The overlap $q$ as function of $s\equiv t/t_{w}$ for $t_w= 2^{9},\ 2^{13}$ and $2^{17}$,
averaged over 3 initial conditions, the errors being  smaller than the points.} \label{Q}
\end{figure}

In order to verify the phenomenon of aging we have we have introduced the 
quantity $q(t_w,t)$ \cite{PAAGE1} defined as
\bea q(t_w,t) \equiv N^{-1}\sum_{i}q_{i}(t_w,t), \\
q_{i}(t_w,t)\equiv \sum_{k} w(x_{i}(t+t_{w})-x_{k}(t_{w})),
\eea
where the sum over $k$ is done over particles of the same type of $i$.  We have chosen the function 
$w$ in such a way that it is very small when $x>>a$ and it is near to $1$ for $x<a$, i.e.  
$w(x)=a^{12}/(x^{12}+a^{12})$, with $a=.22$.  The value of $q$ will thus be a number very near to 
$1$ for similar configurations (in which the particles have moved of less than $a$) and it will be 
much smaller ($O(10^{-1})$) for unrelated configurations; using the terminology of spin glasses $q$ 
can be called the overlap of the two configurations.  In fig.
\ref{Q} we plot the overlap as function of $s\equiv t/t_{w}$ at different values of the waiting
time, 
i.e.$t_w= 2^{9},\ 2^{13}$ and $2^{17}$.  The data weakly depend on the waiting time.  It is 
possible to assume that $q$ goes to limit when $t_{w}\to \infty$ at fixed $s$, this limit being 
reached from above at small $\eps$ and from below at large $\eps$.  This way of approaching the 
asymptotic limit is quite a common feature in other systems, e.g.  spin glasses
\cite{MPROS}.

Our aim is to study the correlations of the particles which have moved in a sizable way when the 
time changes form $t_{w}$ to $s\  t_{w}$.  These particles, unless they have exactly replaced  
other particles, contribute to the decrease of the value of $q$ and have a correlation in in position 
space which strongly depends on waiting time.  To evidenziate this effect \cite{LAPA,KG}, for each 
particle and pairs of configuration at time $t_{w}$ and $s\ t_{w}$ we define the quantity:
\be
\si_{i}(t_w,s\ t_{w})=2(q_{i}(t_w,s\ t_{w})-q(t_w,s\ t_{w})).
\ee
In this way $\sum_{i}\si_{i}=0$.  When the average value of $q$ is  not far 1/2, this procedure 
corresponds to put $\si_{i}\approx 1$, if the particle has moved less than $a$, and $\si_{i}\approx 
-1$, if the particle has moved more than $a$.  Movements which correspond to an interchange of 
particles of the same type have no effect on $\si$.\begin{figure}[htbp]
\epsfxsize=250pt\epsffile{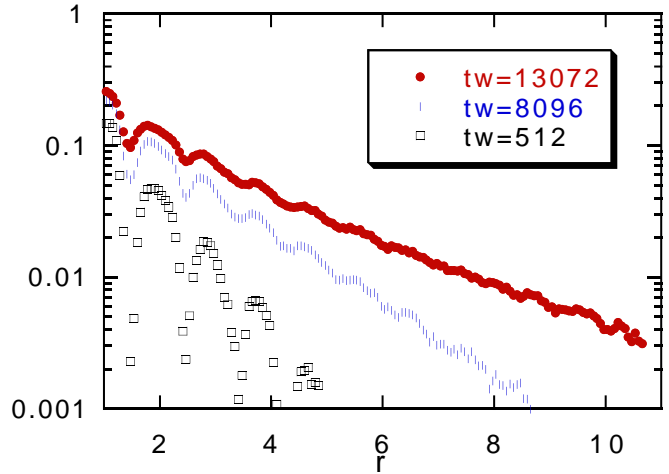}
\caption{The correlation $c(r,t_{w})$ as function of the distance for $t_w= 2^{9},\ 2^{13}$ and 
$2^{17}$ }
\label{F}
\end{figure}
 
We now we define the function $f(r,t_{w})$ as the correlation of the particles at time $s\ t_{w}$, 
where the contribution of two particle is weighted by a factor $\si_{i}\si_{k}$.  Equivalently, we 
can define the function $\mu(x)=\sum_{i}\delta(x-x_{i}(t_{w}))\si_{i}(t_w),$ which is the local 
density $\rho(x)$ multiplied by $\si$.  In this way we can write $ f(r)=\lan \mu(x)\mu(y) \ran, $ 
where $|x-y|=r$.

The correlation $f(r)$ go to zero at large distances, by construction.   We expect also 
that the values of $r$ at which $f(r)$ is sizable different from zero correspond to the values of 
distance of particles which move in a correlated way.

The results for the correlation $f(r,t_{w})$ computed with $s=3$ are shown in fig.
\ref{F}, where we plot $c(r,t_{w})\equiv f(r,t_{w})/g(r)$ (we have divided the data for 
$f(r,t_{w})$ 
by $g(r)$ in order to eliminate a natural oscillatory effect).  At short times the correlations are 
present only at short distances, when the time increases they extend on a much larger region.  We 
have analyzed the data by fitting $c(r,t_{w})$ as
\be
c(r,t_{w})=\exp(-r/R(t_{w}))(c_{1}+c_{2}r^{-1}{\sin(2 \pi r/d+\phi)}),
\ee
where we have taken the value of $d$ from the previous fit of the correlation function $g(r)$.  
(The 
same qualitative dependence of the dynamic correlation distance $R(t_{w})$ is obtained also using 
other fitting procedures; the precise value of $R(t_{w})$ being also here quite sensitive to the 
introduction of a power decaying prefactor).

\begin{figure}[htbp]
 \epsfxsize=250pt\epsffile{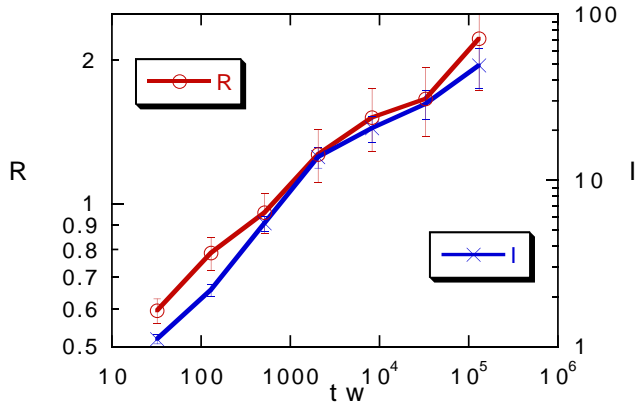}
\caption{On a double logarithmic scale we find the values of the correlation functions ($R(t_{w})$,
axis
on 
the left) as function of the waiting time and the values of the integral of the correlation function 
($I(t_{w})$, axis on the right).}
\label{I}
\end{figure}

In fig.  \ref{I} we show the dynamic correlation distance as function of time.  We see that the 
behaviour of $\R(t_{w})$ is  well approximated by a power of time, the exponent being about 
0.16.  

An other interesting quantity is $I(t_{w})\equiv \int d^{3}x f(x,t_{w})$ that 
is the equivalent of a susceptibility (it is equal to $N (\lan q(t_{w},s\ t_{w})^{2}\ran - \lan 
q(t_{w},s\ t_{w})\ran^{2}$).  Roughly speaking it is proportional to the number of particles which 
move in a correlated way.  If the typical event correspond to the formation of a fluidified domain 
of radius of order $R(t_{w})$, $I(t_{w})$ should be proportional to $R(t_{w})^{3}$.  The corresponding 
exponent for $I$ is about $0.52$ (a similar conclusion was reached in \cite{PAAGE1} by an analysis 
restricted to much smaller samples, $N\le 258$).  It would be tempting to speculate that the exact 
exponents are $1/6$ and $1/2$.

We have also similar data also for different values of $s$, i.e.  $s=1.5$ and $s=.375$.  The 
dependence on $s$ of $R(t_{w})$ is  mild and the analysis of these data leads to similar 
conclusions.  The data for $f(r,t_{w})$ at different $s$ are indeed rather similar and there is 
mainly a difference in the absolute normalization.  This fact is consistent with the possibility 
that the correlated movements of a large number of particles happen in a time interval which is 
much smaller of $t_{w}$ (this point should be studied more carefully).

Our data indicate that the process of rearrangements of the systems happens on a spatial scale 
which becomes larger and larger as function of the time.  Events that correspond to the 
rearrangements of regions of size $R$ have a characteristic time which diverge approximately as 
$R^{6}$.  A comparison with the data of \cite{BAPA} shows that in our simulations we have explored 
a 
region of time where  the relaxation of energy indicates that the approach to 
equilibrium is dominated by crossing of entropic barriers.  For values of time near to the largest 
ones used in our simulations the energy relaxation enter in a new regime
\cite{BAPA}, which is quite likely dominated by energetic barriers.  It would be extremely 
interesting to extend these studies to longer times and larger systems, to see if there is any 
change in the behaviour of $R(t_{w})$ when we enter in this new regime.

I thank Walter Kob and David Lancaster for useful discussions.

\end{document}